# Efficiency of genetic algorithm and determination of ground state energy of impurity in a spherical quantum dot

**Haluk Şafak[1*], Mehmet Şahin[1], Berna Gülveren[1], Mehmet Tomak[2]**

[1]Selcuk University, Faculty of Arts and Science, Physics Dept. Kampus 42075, Konya, Turkey
[2]Middle East Technical University, Physics Dept. 06531, Ankara, Turkey

## Abstract

In the present work, genetic algorithm method (GA) is applied to the problem of impurity at the center of a spherical quantum dot for infinite confining potential case. For this purpose, any trial variational wave function is considered for the ground state and energy values are calculated. In applying the GA to the problem under investigation, two different approaches were followed. Furthermore, a standard variational procedure is also performed to determine the energy eigenvalues. The results obtained by all methods are found in satisfactory agreement with each other and also with the exact values in literature. But, it is found that the values obtained by genetic algorithm based upon wavefunction optimization are closer to the exact values than standard variational and also than genetic algorithm based on parameter optimization methods.

## 1. Introduction

The studies performed so far to determine the electronic properties of semiconductor quantum nanostructures are based generally either on variational or on numerical procedures [1-3]. However, recently a somewhat different optimization technique, the so-called evolutionary genetic algorithm (GA), has appeared to be used more frequently in the optimization and minimization problems for the quantum mechanical systems [5-7]. This method, since firstly proposed by Holland [8], has been applied to many scientific areas, for instance, in engineering optimization and improvement problems, for energy calculations in atomic and molecular physics, in nonlinear fitting problems and at the crystal growth studies in condensed matter [9-13]. It relies essentially on variational procedure. But GA exhibits some important differences from other traditional variational optimization procedures. These

---

**e-mail:** hsafak@selcuk.edu.tr

differences can be summarized as follows: (i) GA employs the coding of any parameter (or parameter set), not parameters themselves, (ii) it starts from any initial population of possible solution, not from a single value or analytical expression, (iii) it uses some fitness (or objective) information in procedure, not any derivative or auxiliary knowledge, and (iv) it follows the probabilistic rules, instead of the deterministic ones.

Although the operations performed in any GA procedure may differ in practice, the main purpose is always the same, namely to select the fittest member(s) of any chosen initial population of possible solutions. In other words, a genetic algorithm allows any initial population consisting of many individuals to evolve, under some pre-specified selection rules, toward a state that maximizes the fitness. To this end, GA keeps on reproducing good member(s) while killing the bad ones. Therefore, any selection process is implemented to eliminate the low-fitness individuals from the population with more probability and then to let survive the good-fitness ones and copy them to the next generations. This selection procedure is performed regarding to the fitness values of individuals.

In the actual application of the GA procedure to any quantum-mechanical system, an initial trial ground state wave function set is firstly chosen according to the specific problem and then, energy values are calculated by using this function set. Next, GA begins with these initial eigenvalues and performs some genetic operations over them. These operations and flow course of algorithm are very different than other traditional optimization methods.

In this paper, we apply the evolutionary GA procedure to the problem of impurity located at the center of a spherical dot with infinite confining potential. In addition to GA method, we have also performed the standard variational calculations. Energy values are calculated only for the ground state. In applying the GA procedure to the problem, we have followed two different approaches. In one of them, we have optimized the variational parameter in minimization process. But, in the other, the wave function itself has been selected as the parameter to be optimized.

The main purpose of this work is to check the efficiency of GA against the standard variational method, and also to compare two different GA based optimization methods in according to their applicability and performance in attainment the accurate results in such a problem.

## 2. Theory

It is well known that the Hamiltonian for an on-center impurity in a spherical quantum dot can be written, for the ground state, as

$$H = -\frac{\hbar^2}{2m}\nabla_r^2 - \frac{e^2}{r} + V(r) \qquad (1)$$

where

$$\nabla_r^2 = \frac{d^2}{dr^2} + \frac{2}{r}\frac{d}{dr} \qquad (2)$$

is the Laplacian in the spherical coordinates, $V(r)$ is the confining potential, e is the electronic charge and m is the electron mass. The solution of Schrödinger equation for a single hydrogenic impurity would then require the solution of an energy eigenvalue problem analytically or numerically. Different approaches may be made to solve the Schrödinger equation, e.g. variational methods [1,2], perturbative approximations [3] or numerical procedures [4]. In every approach, it is desired to find a suitable $\psi_n$ wave function describing the system adequately well and next to obtain scalar $E_n$ eigenenergies which are the solutions of the following set of homogeneous differential equations under appropriate boundary conditions

$$(H - E_n)\psi_n = 0 \qquad (3)$$

In this work, our aim is to determine the eigenenergies for an on-center impurity in a spherical quantum dot confined by an infinite potential well. Here we consider only the ground state case. Therefore, we have chosen the wave function describing the system as,

$$\psi(r) = \begin{cases} A\dfrac{\sin(kr)}{r}\exp(-\lambda r), & r < R \\ 0, & r > R \end{cases} \qquad (4)$$

which should satisfy the boundary conditions, e.g. must vanish at the spherical dot boundary, r=R, R is being the dot radius. Here k is $\pi/R$ and $\lambda$ is variational parameter. The amplitude A is then calculated by the normalization procedure,

$$\int \psi^*(r)\psi(r)d^3r = 1 \qquad (5)$$

for each value of the λ parameter. The V(r) confining potential can be defined as

$$V(r) = \begin{cases} 0 & r < R \\ \infty & r > R \end{cases} \qquad (6)$$

## 3. Method

In this study, we follow three different approaches to determine the ground state energy of hydrogenic impurity. One of them is the standard variational method and two others are based on evolutionary genetic algorithm.

### 3.1 Standard variational procedure

As known, the standard variational procedures are based essentially on the energy minimization requirement. In this method, any trial wave function describing the system is chosen, including one (or more) variational parameter(s). Next, using this wave function, energy eigenvalues are calculated numerically from energy minimization condition stated as

$$E = \min_{\lambda} \frac{\left\langle \psi^*(\lambda, r) \middle| H \middle| \psi(\lambda, r) \right\rangle}{\left\langle \psi^*(\lambda, r) \middle| \psi(\lambda, r) \right\rangle}. \qquad (7)$$

The wave function (and therefore the λ variational parameter) corresponding to the minimum energy value represents the solution to the problem.

### 3.2 Genetic Algorithm procedure

Genetic algorithms (GA) are general search and numerical optimization algorithms inspired by natural selection processes. As the first step, any initial population for solution is created using the equations describing the system. For each individual at this initial population, employing some systematical requirements (e.g. energy minimization), their fitness values are determined. Next, starting from this initial possible solution population, it is aimed to obtain the optimum solution. For this purpose, some genetic operations are performed.

There are three main operations in GA, which has been employed in many practical applications and yielded satisfactory results. These are, (i) reproduction (or copy), (ii) crossover and (iii) mutation.

Reproduction is a process in which the individuals of any population are copied to next generations, according to their fitness values. Therefore, in performing reproduction process, any selection procedure is necessary to choose the individuals to be copied to the next generations. In implementing the selection, generally both the statistical considerations and the optimization (minimization or maximization) requirements should be regarded.

The crossover operation is performed between two individuals. Two random members are selected from present population. Each member of selected pair is cut at any random crossing site. Next, the components of two individuals are interchanged and so two new individuals are created.

Another operation in GA, the mutation process plays a less important role and is implemented at lower probabilities than other operations. In this process, the genetic information is changed randomly.

Mutation is especially important in cases where to fall into undesired local minima, during the search of global one, has become more probable. But, the mutation rates are usually chosen very small. The selection of high mutation probabilities would destroy the convergence behavior of the numerical minimization process.

At each step of genetic procedure, reproduction, crossover and mutation processes are randomly performed and whole procedure is repeated until a satisfactory convergence is achieved and the optimum solution is found.

In applying genetic algorithm procedure to any physical problem, it is necessary to decide which parameters would be optimized. In this work, we apply the GA procedure in two different ways to the problem of hydrogenic impurity.

## I. GA based on parameter-optimization (GAPO)

In this method, the $\lambda$ variational parameter at Eq.(4) is selected as an individual upon which genetic operations will be performed. So, we have converted $\lambda$ parameter into the binary code as usual way in most genetic algorithm processes. We have chosen the population number (npop) as 100. Thus, an initial population has been created from Eq.(4) for randomly selected values of $\lambda$. By using this initial generation, expectation values of energy are determined from Eq.(7). Then, the fitness values for each member of population are created from these energy values by means of

$$\text{Fitness}\left[\psi_i\right] = \exp\left[-\beta\left(E_i - \left|E_{av}\right|\right)\right], \tag{8}$$

where $\beta$ is a constant and $E_{av}$ is the average of energy eigenvalues. Using this fitness values, a rulet wheel [9] is constituted and a selection procedure has been performed. In this selection procedure, generally, better individuals are selected, however sometimes less fit individuals can also be selected and then new generation is created from this set of chosen individuals. This process is called as reproduction or copy.

In the crossover, two random individuals (or $\lambda$ parameters here) are selected from the population. Each member of selected strings is cut at any random crossing site. After this process, the components of the strings of two individuals interchanged and two new strings are created as illustrated in the following example:

$$\lambda_1: 10101\mathbf{111} = 0.68359375 \qquad \lambda_1^{'}: 10101\mathbf{010} = 0.66406250$$

$$\lambda_2: 10010\mathbf{010} = 0.57031250 \qquad \lambda_2^{'}: 10010\mathbf{111} = 0.58984375$$

In the mutation process, the genetic information (i.e. a bit of any selected individual) is changed randomly. A simple mutation operation can be illustrated as follows:

$$\lambda_1: 101\mathbf{0}1111 = 0.68359375 \qquad \lambda_1^{'}: 101\mathbf{1}1111 = 0.74609375$$

Mutation process is especially important in cases where to fall into undesired local minima during the search of global one has become more probable. But the mutation rates are usually chosen very small. Because, the selection of high mutation probabilities would destroy the convergence behavior of numerical minimization process and therefore give rise to ambiguities.

## II. GA based on wavefunction-optimization (GAWO)

In this method, we select the $\psi(r)$ wave function (not variational parameter) as the genetic code, since all information necessary for specifying the system has been stored it, and all calculations are performed numerically by employing this wave function. Firstly, the initial population has been created numerically from Eq.(4) for randomly values of $\lambda_k$ (k=1...npop) and assigned to two dimensional vector arrays. This population has been normalized numerically by using Eq.(5). Thus, a normalized random population of wave functions (individuals) is created as an initial generation. Eigenenergy values are determined from this generation by means of

$$E_k = \left\langle \psi_k^*(r) \middle| \hat{H} \middle| \psi_k(r) \right\rangle \qquad (9)$$

Fitness values are created from these energy values again by using Eq.(8). Rulet wheel is constituted and the selection procedure has been performed.

In the crossover, we take two randomly chosen individuals (or wave functions). Two new functions are produced by using these individuals as

$$\psi_1^{'}(r) = \psi_1(r)S(r) + \psi_2(r)[1 - S(r)]$$

$$\psi_2^{'}(r) = \psi_2(r)S(r) + \psi_1(r)[1 - S(r)] \qquad (10)$$

here S(r) is a smooth step-function, which can be defined as follows,

$$S(r) = \frac{1}{2}\left[1 + \tanh\left((r - r_0)/w^2\right)\right] \qquad (11)$$

with $r_0$ is again any random parameter chosen at interval (0,R), R being the dot radius, and w is a parameter which defines the sharpness of crossover process [7]. Once these two new functions are created, then the fitness values corresponding to them are determined. The probability for crossover operation in each GA step was chosen as $P_c$=0.95.

In performing the mutation operation, we have randomly selected from initial population any $\psi_M(r)$ sample mutation function as corresponding to any random $\lambda$ parameter and added this function to any other randomly chosen $\psi(r)$ function to create a new parent function as follows,

$$\psi'(r) = \psi(r) + \psi_M(r) . \qquad (12)$$

We have selected the mutation probability as small as possible, e.g. $P_m$=0.05.

In each step of GA procedure, the population sizes are kept constant and equal npop=100, and any selection operation among individuals are performed to reproduce (or copy) those with high fitness values to next generations, while to eliminate the bad-fitness ones.

## 4. Results and Discussion

The results obtained are listed in Table 1 together with the exact values [3,14]. We have used atomic units at all calculations, where $\hbar = 1$, the electronic charge e=1 and the electron mass m=1. It can be seen from Table 1 that the results obtained by GAWO are generally closer to the exact values than both the variational and also the GAPO results. For both positive and negative energy values, same good accuracy is achieved.

In Table 2, the values for variational parameter $\lambda$ obtained by standard variational and GAPO methods are listed. As seen from results, the variational and GAPO values are very close to each other at almost each dot radii. Therefore, the applying genetic algorithm for optimization the parameter has not given any significantly different results than the variational procedure.

In figure 1, we have plotted two normalized wave functions $\psi(r)$ as a function of r, one of them (dashed curve) is obtained variationally and the other (solid line) is by GAWO procedure after all iterations. As shown from the figure, the both wave functions vanish at the spherical dot boundary. But, at r=0 a marked difference has been appeared between two wave functions. The curve obtained by GAWO is starting from larger values than the variational wave function (and also than one obtained by GAPO). The wave function used in variational (and also in GAPO) method remains in nearly same form during all the calculation procedure, only variational parameter has been changed. Therefore, at near r=0, the value of wavefunction has been approached to the normalization constant as expected. However, in the genetic algorithm method based-wavefunction optimization, the wavefunction in its analytical form is used only for construction the initial population, and then, all processes have been performed on the numerical values of function, not on its analytical form. Thus, in the applying the procedure, the analytical form of initial wavefunction has not any significant effect on the later numerical calculations, instead the energy minimization requirement has become dominant factor at the rest of procedure. Therefore, it is not meaningless result that the wavefunction approaches toward a somewhat different value from its starting analytical form.

It is seen in all calculations that genetic algorithm procedure is very efficient in obtaining the minimum value of energy. It reaches this minimum in a surprisingly short time in comparison to the direct variational method. For example, the convergence here is achieved only after a couple of ten iterations, and it is seen also that these results obtained have been found very reproducible in many several times trials. In figure 2, it is given the evolution of energy eigenvalue for the dot radius R=1 with the number of iterations. As seen from figure, the energy eigenvalue converges to a constant value after nearly 20 iterations.

## 5. Conclusion

The standard variational procedure has been used over many years in obtaining acceptable solutions to the scientific problems, especially if any analytical solution can not be derived easily. However, if the number of variational parameters is more than one, then it would be more difficult to perform direct variational procedure with the same accuracy and easiness. This method is very sensitive to the starting value and the range of the selected variational parameter. But, these kinds of restrictions are less important in GA process. This advantage gives rise to an increase in the popularity of GA over other techniques.

GA works with the fitness values which are based mainly on the predetermined initial population for solutions and employs some random probabilistic transition procedure, not deterministic rules, in attaining to the desired optimal point, and therefore the procedure is less sensitive to the starting point or the complexity of the problem comparison to the other traditional variational methods. But, in some problems, which contain more complexity, this method can also get stuck.

Here, we apply the GA in two different ways as explained in the above. The one in which we optimized the variational parameter by genetic algorithm has not given considerably different values than the standard variational calculations. But, the wavefunction optimization procedure has provided more accurate results. So, it can be concluded that it is more convenient to perform the GA for optimization the wavefunction itself, not the variational parameter in such a problem.

As far as the quantum dot problem is concerned, we believe the GA method to be a powerful alternative to other computational tools. This has been recently demonstrated for the case of a few electrons in a quantum box [15].

**References**
1.  J. Bellessa, M. Combescot, Solid State Comm., **111**, (1999) 275
2.  Y. P. Varshni, Phys. Lett. A, **252** (1999) 248
3.  C.Bose and C.K. Sarkar, Phys. Stat. Sol.(b), **218**, (2000) 461
4.  K.F. Ilaiwi and M. Tomak, Solid State Comm., **78**, (1991) 1007

5. P. Chaudhury, S. P. Bhattacharyya, Chem. Phys. Lett. **296** (1998) 51

6. H. Nakanishi, M. Sugawara, Chem. Phys. Lett. **327** (2000) 429

7. I. Grigorenko, M. E. Garcia, Physica A, **284** (2000) 131

8. J. H. Holland, Adaptation in Natural and Artifical Systems, University of Michigan Press, (1975)

9. D. E. Goldberg, Genetic Algorithms in Search, Optimization, and Machine Learning, Addison-Wesley, Reading, MA, (1999)

10. R. Saha, P. Chaudhury, S. P. Bhattacharyya, Physics Lett. A, **291** (2001) 397

11. L. Liu, L. Zhao, Y. Mao, D. Yu, J. Xu, Y. Li, Int. J. Mod. Phys. C **11** (2000) 183

12. A. Brunetti, Comp. Phys. Comm. **124** (2000) 204

13. Ö. Şahin, P. Sayan, A. N. Bulutcu, J. Crys. Growth **216** (2000) 475

14. J. L. Marin, S. A. Cruz, Am. J. Phys. **59** (1991) 931

15. I. Grigorenko, M. E. Garcia, Physica A, **291** (2001) 439

**Table 1.**

The ground state energy values as a function of radius R of a spherical dot, obtained by three different methods. The agreement between the exact and GAWO results is quite good. Energies are expressed in Rydbergs and radii in Bohr units.

| R | Energy (Exact) | Energy (Variational) | Energy (GAPO) | Energy (GAWO) |
|---|---|---|---|---|
| 0.5 | 29.4959 | 29.5241 | 29.5241 | 29.4984 |
| 1.0 | 4.7480 | 4.7768 | 4.7767 | 4.7517 |
| 1.2 | 2.5386 | 2.5675 | 2.5674 | 2.5401 |
| 1.6 | 0.5426 | 0.5715 | 0.5714 | 0.5445 |
| 1.8 | 0.0651 | 0.0938 | 0.0938 | 0.0662 |
| 2.0 | -0.2500 | -0.2216 | -0.2216 | -0.2489 |
| 2.4 | -0.6128 | -0.5854 | -0.5854 | -0.6116 |
| 2.8 | -0.7933 | -0.7675 | -0.7676 | -0.7915 |
| 3.0 | -0.8479 | -0.8231 | -0.8231 | -0.8446 |
| 3.2 | -0.8878 | -0.8641 | -0.8640 | -0.8855 |
| 3.6 | -0.9387 | -0.9175 | -0.9175 | -0.9361 |
| 3.8 | -0.9547 | -0.9349 | 0.9348 | -0.9540 |
| 4.0 | -0.9665 | -0.9481 | -0.9481 | -0.9641 |
| 5.0 | -0.9928 | -0.9811 | -0.9811 | -0.9922 |
| 6.0 | -0.9985 | -0.9918 | -0.9917 | -0.9966 |
| 7.0 | -0.9997 | -0.9958 | -0.9958 | -0.9992 |

**Table 2.**

The values for variational parameter $\lambda$ obtained by standard variational and GAPO methods. As seen from table, the values are very close to each other at almost each dot radii.

| R | $\lambda$ (Variational) | $\lambda$ (GAPO) |
|---|---|---|
| 0.5 | 0.4660 | 0.4648 |
| 1.0 | 0.4960 | 0.4961 |
| 2.0 | 0.5740 | 0.5742 |
| 3.0 | 0.6610 | 0.6602 |
| 4.0 | 0.7510 | 0.7500 |
| 5.0 | 0.8200 | 0.8203 |
| 6.0 | 0.8680 | 0.8670 |
| 7.0 | 0.9010 | 0.9023 |

**Figure Caption**

**Figure 1:** The variation of wave functions, obtained by standard variational and GAWO methods, as a function of dot radii. Dashed curve corresponds to the variational and solid line to one obtained by GAWO.

**Figure 2:** The evolution of energy eigenvalue for R=1 dot radius with the number of iteration.

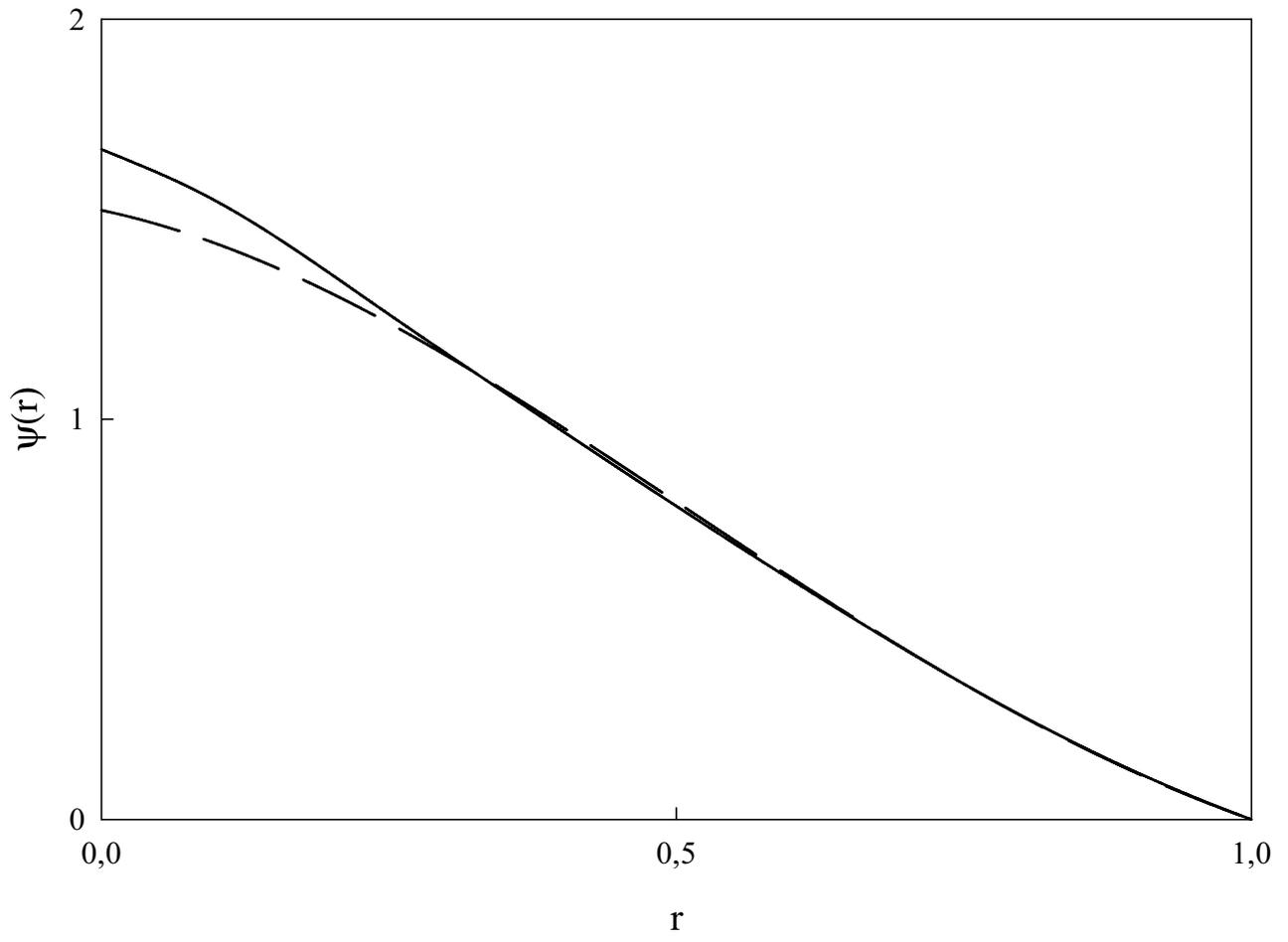

**Figure 1**

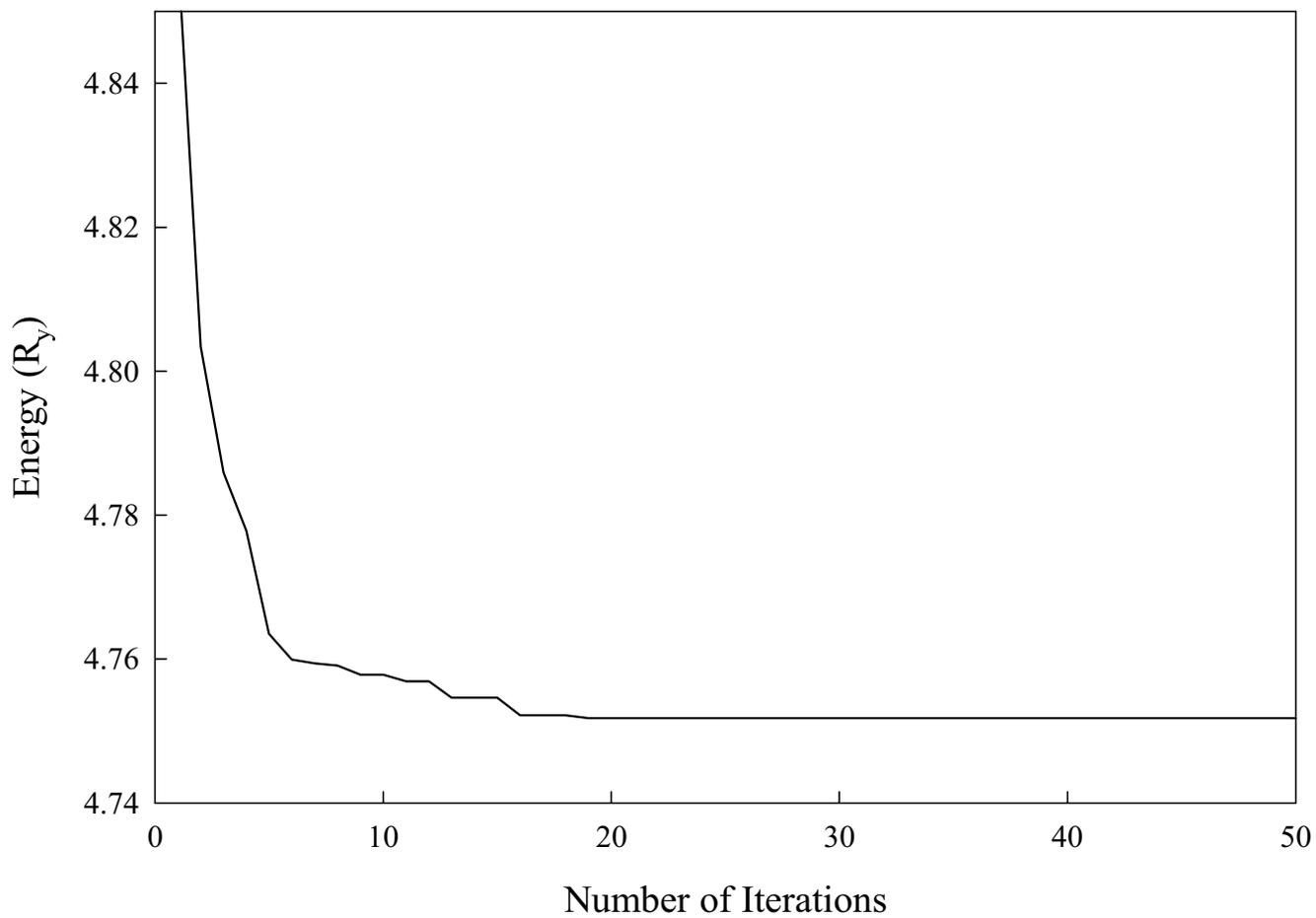

**Figure 2**